\documentstyle[12pt]{amsart}
\newtheorem{Lemma}{Lemma}
\newtheorem{Proposition}{Proposition}
\newtheorem{Theorem}{Theorem}
\newtheorem{Corollary}{Corollary}
\def\proof{\par{\it Proof}. \ignorespaces}
\def\endproof{{\ \vbox{\hrule\hbox{%
   \vrule height1.3ex\hskip0.8ex\vrule}\hrule }}\par}
\newenvironment{Proof}{\proof}{\endproof}

\begin{document}

\title{Iso-spectral deformations of general matrix and \\
their reductions on Lie algebras}

\author{Yuji KODAMA$^*$ \ \
 and  \ Jian YE$^\dagger$ \\ }

\thanks{*Department of Mathematics, The Ohio State University,
Columbus, OH 43210\endgraf
{\it E-mail address\/}: kodama@@math.ohio-state.edu
\endgraf$\dagger$ Department of Mathematics, The Ohio State University,
Columbus, OH 43210\endgraf
{\it E-mail address\/}: ye@@math.ohio-state.edu}

\date{(June, 1995)}
\maketitle

\begin{abstract}

We study an iso-spectral deformation of general matrix which is
a natural generalization of the Toda lattice equation.
We prove the integrability of the deformation, and
give an explicit formula for the solution to the initial value problem.
The formula is obtained by generalizing
the orthogonalization procedure of Szeg\"{o}.
Based on the root spaces for simple Lie algebras, we consider several
reductions of the hierarchy.  These include not only the integrable systems
studied by Bogoyavlensky and Kostant, but also their generalizations
which were not known to be integrable before.
The behaviors of the solutions are also studied. Generically, there
are two types of solutions, having either sorting property or blowing up
to infinity in finite time.
\end{abstract}

\medskip

{\bf Mathematics Subject Classifications (1991).} 58F07, 34A05

\par\bigskip\bigskip

%\newline\newline

\section{Introduction}
\renewcommand{\theequation}{1.\arabic{equation}}\setcounter{equation}{0}

In this paper we consider an iso-spectral deformation of an arbitrary
diagonalizable matrix $L \in {\frak M}(N,{\Bbb R})$.  With the deformation
parameter $t \in {\Bbb R}$, this is defined by
\begin{eqnarray}
\label{gtoda}
\frac{d}{dt} L = \left[ P \ , \ L \right] \ ,
\end{eqnarray}
where $P$ is the generating matrix of the deformation, and is given by
a projection of $L$,
\begin{eqnarray}
\label{pmx}
P = \Pi (L) := \left( L \right)_{>0} -
\left( L \right)_{<0} \ .
\end{eqnarray}
Here $\left( L \right)_{>0 \ (<0)}$ denotes the strictly upper (lower)
triangular part of $L$.  In terms of the standard basis of ${\frak M}(
N,{\Bbb R})$, i.e.
\begin{equation}
\label{e}
E_{ij} = \left( \delta_{ik}\delta_{j\ell} \right)_{1 \le k, \ell \le N} \ ,
\end{equation}
the matrices $L$ and $P$ are expressed as
\begin{eqnarray}
\label{l}
L &=& \sum_{1 \le i, j \le N} a_{ij} E_{ij} \ , \\
\label{p}
P &=& \sum_{1 \le i < j \le N} a_{ij} E_{ij} \ - \
\sum_{1 \le j < i \le N} a_{ij} E_{ij} \ .
\end{eqnarray}
Using the commutation relations for $E_{ij}$, i.e.
\begin{equation}
\label{comm}
[ E_{ij} \ , \ E_{k\ell}] = E_{i\ell}\delta_{jk} - E_{jk}\delta_{i\ell} \ ,
\end{equation}
the equations for the components $a_{ij}=a_{ij}(t)$ are written in the form,
\begin{eqnarray}
\label{ctoda}
\nonumber
\frac {d a_{ij}}{dt}  &=& 2 \left( \sum_{k=I+1}^{N} - \sum_{k=1}^{J-1} \
 \right) a_{ik}a_{kj} \\
& & + \left(a_{II} - a_{JJ} \right) a_{ij} \ ,
\end{eqnarray}
where $I:= max(i,j)$ and $J:=min(i,j)$.  The equation (\ref{gtoda}) is
also defined as the compatibility of the following linear equations with
iso-spectral property of $L$,
\begin{eqnarray}
\label{leq}
L \Phi &=& \Phi \Lambda \ , \\
\label{peq}
\frac {d}{dt} \Phi &=&  P \Phi \ ,
\end{eqnarray}
where $\Phi$ is the eigenmatrix, and $\Lambda$ is the diagonal matrix of
eigenvalues,
\begin{equation}
\label{ev}
\Lambda = diag(\lambda_1, \cdots, \lambda_N) \ .
\end{equation}
The set of equations (\ref{leq}) and (\ref{peq}) is also referred as the
inverse scattering transform for the system (\ref{gtoda}).  The main
purpose of this paper is to show the complete integrability of (\ref{gtoda})
with (\ref{l}) and (\ref{p}) by means of the method of inverse scattering
transform.

\medskip

One of the most famous and important example of (\ref{gtoda}) is
of course the non-periodic Toda lattice equation, where $L$ is given by a
symmetric tridiagonal matrix \cite{Mo}.  The matrices $L$ and $P$ for this
equation are commonly written as
\begin{eqnarray}
\label{todal}
L_{T} &=& \sum_{i=1}^{N} a_i E_{ij} \ + \
\sum_{i=1}^{N-1} b_i \left( E_{i,i+1} + E_{i+1,i} \right) \ , \\
\label{todap}
P_{T} &=& \sum_{i=1}^{N-1} b_i \left( E_{i,i+1} - E_{i+1,i} \right) \ .
\end{eqnarray}
The integrability of the Toda lattice equation has been shown by the inverse
scattering method \cite{F} \cite{Ma} \cite{Mo}.
In this paper, we call (\ref{gtoda})
with (\ref{pmx}) the ``generalized Toda equation".

\medskip

Several generalizations of the Toda lattice equation have been considered.
In \cite{B}, Bogoyavlensky extended the equation based on the simple
roots of semi-simple Lie algebra $\frak g$, where $L$ and $P$ were given by
\begin{eqnarray}
\label{bogl}
L_B &=& \sum_{i=1}^r a_i h_i \ + \ \sum_{\alpha \in \Pi} b_{\alpha}
(e_{\alpha} + e_{-\alpha}) \ , \\
\label{bogp}
P_B &=& \sum_{\alpha \in \Pi} b_{\alpha} (e_{\alpha} - e_{-\alpha}) .
\end{eqnarray}
Here the elements $h_i, e_{\alpha}, e_{-\alpha}$ are Cartan-Weyl bases in
$\frak g$ with $r=rank (\frak g)$ and $\Pi=$ the set of the simple roots.
All of these equations associated with semi-simple Lie algebras are shown
to be completely integrable hamiltonian systems.
In \cite{Ko} Kostant solved these by using the representation theory
of semi-simple Lie algebras.
In \cite{BBR}, Bloch et al. showed that these systems can be also written
as gradient flow equations on an adjoint orbit of  compact Lie group.
They then showed that the generic flow assumes the ``sorting property"
(or convexity).
Here the sorting property means that
$L(t) \rightarrow
\Lambda=diag\left( \lambda_1, \cdots, \lambda_N \right)$ as $t\rightarrow
\infty$, with the eigenvalues being ordered by $\lambda_1>\lambda_2>\cdots>
\lambda_N$.

\medskip

There are also other types of extensions:  One of them is to extend $L_{T}$
in (\ref{todal}) to a full symmetric matrix.  The corresponding system,
which we call the ``full symmetric Toda equation", was shown by Deift et al.
\cite{DLNT} to be also a complete integrable hamiltonian system.  In \cite{KM}
Kodama and McLaughlin solved the initial value problem of the corresponding
inverse scattering problem (\ref{leq}) and (\ref{peq}), and obtained
an explicite formula of the solution in a determinant form.  They also showed
the sorting property in the generic solution.  As a slight generalization of
the full symmetric Toda equation, Kodama and Ye \cite{KY} considered a system
with symmetrizable matrix $L$, which is expressed as $L_{KY}=L_SS$ with
a full symmetric matrix $L_S$ and a diagonal matrix $S$.  A key feature of
this system is that the eigenmatrix of $L_{KY}$ can be taken as an element of
noncompact group of matrices, such as $O(p,q)$, and defines an indefinite
metric in the eigenspace.  The integrability was also shown by a similar way
as in \cite{KM}, and the general solution now assumes either sorting property
or blowing up to infinity in finite time as a result of the indefinite metric.

\medskip

In \cite{EFS}, Ercolani et al. proposed the equation (\ref{gtoda})
with matrices,
\begin{eqnarray}
\label{hl}
L_H &=& \sum_{i=1}^{N-1} E_{i,i+1} \ + \ \sum_{1 \le j \le i \le N}
b_{ij} E_{ij} \ , \\
\label{hp}
P_H &=& -2 (L_H)_{<0}\ =  \ - 2\sum_{1 \le j < i \le N} b_{ij} E_{ij} \ ,
\end{eqnarray}
which was called the ``full Kostant-Toda equation".  This is also an extention
of the Toda equation (\ref{todal}) which can be written in the form,
\begin{equation}
\label{ttodal}
{\tilde L}_{T} = \sum_{i=1}^{N-1} E_{i,i+1} \ +\ \sum_{i=1}^N a_i E_{ii} \
+ \ \sum_{i=1}^{N-1} b_i^2 E_{i+1,i} \ .
\end{equation}
As we will show in this paper, the transformation from (\ref{todal})
to (\ref{ttodal}) is given  by a rescaling of the eigenvectors of $L_T$.
Several group theoretical structure of the extended system were found.  However
the question whether the system is completely integrable still remains
open in a sense of explicit integration.

\medskip

It is immediate but important to observe that all of these extensions are
special reductions of the generalized Toda equation (\ref{gtoda}).
In fact, we show that
these reductions are obtained more systemalically as crtain decompositions
of the root spaces of semi-simple Lie algebras.

\medskip

In this paper we prove the complete integrability of any reductions
of the generalized Toda equation (\ref{gtoda}).  The content of this paper
is as follows:   We start with a preliminary
in Section 2 to give some background information necessary for analysis
of the system (\ref{gtoda})
and the inverse scattering scheme (\ref{leq}) and (\ref{peq}).

\medskip

In Section 3, we solve the initial value problem of (\ref{peq})
for the general system (\ref{gtoda}) by
generalizing the method developed in \cite{KM} and \cite{KY}.
A key in the method is to use  the orthonormalization
procedure of Szeg\"o, which is equivalent to the Gram-Schmidt orthogonalization
method.  This shows the complete integrability of the generalized Toda
equation in the sense of inverse scattering transformation method.

\medskip

In Section 4, we present reductions of (\ref{gtoda}) according to
the classification of semi-simple Lie algebras.  The matrix $L$ here then
contains ``all" the root vectorts, and it gives a generalization of the
system formulated by Bogoyavlensky \cite{B}.  A key ingredient here is to
find a matrix representation of the algebra in a decomposition consisting
of diagonal, strictly upper and lower matrices (Lie's Theorem \cite{H}).
Then the integrability of
these systems associated with semi-simple Lie algebras is a direct consequence
of the result in Section 3.

\medskip

Section 5 provides other reductions which include the full Kostant-Toda
equation and system with a matrix $L$ having band structure in the
elements.

\medskip

In Section 6, we discuss the behavior of the solutions.  Generically,
in addition to the sorting property, there are slutions blowing up
to infinity in finite time, as in the case discussed in \cite{KY}.

\medskip

Finally we illustrate the  results obtained in this paper
with explicit examples in Section 7.

\section{Preliminary}
\renewcommand{\theequation}{2.\arabic{equation}}\setcounter{equation}{0}

Here we give some background information necessary for the inverse scattering
method (\ref{leq}) and (\ref{peq}).  As we will see in the next section,
a key idea for solving these equations is to use an orthogonality of the
eigenfunctions of (\ref{leq}).  This is simply to consider a
dual system of (\ref{leq}) and (\ref{peq}), which are written by
\begin{eqnarray}
\label{dleq}
 \Psi { L} &=& \Lambda \Psi \ , \\
\label{dpeq}
\frac{d}{dt} \Psi &=& - \  \Psi  P \ ,
\end{eqnarray}
where the matrix $\Psi$ is taken to be $\Phi^{-1}$, and of course
\begin{equation}
\label{ids}
\Psi \Phi = I , \, \, \, \ \
\Phi \Psi = I \ .
\end{equation}
In terms of the eigenvectors,
these matrices can be expressed as
\begin{eqnarray}
\label{phi}
& &\Phi \equiv \left[ \phi(\lambda_{1}), \ \cdots \ ,
\phi(\lambda_{N}) \right] \ =
\ \left[ \phi_{i}(\lambda_{j}) \right]_{1 \le i , j \le N} \ . \\
\label{psi}
& & \Psi \equiv \left[\psi(\lambda_1), \ \cdots \ ,
\psi(\lambda_N) \right]^T \
= \ \left[\psi_j(\lambda_i) \right]_{1 \le i, j \le N} \ .
\end{eqnarray}
Note here that $\phi(\lambda_i)$ and $\psi(\lambda_i)$
are the column and row eigenvectors, respectively.
Then the equations (\ref{ids}) give
\begin{eqnarray}
\label{ortho1}
\sum_{k=1}^{N} \psi_{k}(\lambda_{i}) \phi_{k}(\lambda_{j}) &=&
\delta_{ij}   \ , \\
\label{ortho2}
\sum_{k=1}^{N} \phi_{i}(\lambda_{k}) \psi_{j}(\lambda_{k}) &=&
\delta_{ij}   \ ,
\end{eqnarray}
which are called the ``first and second orthogonality conditions".
With (\ref{ortho2}), one can define an inner product $<\cdot,\cdot>$ for
two functions $f$ and $g$ of $\lambda$ as
\begin{eqnarray}
\label{product}
<f, g> := \sum_{k=1}^Nf(\lambda_k)g(\lambda_k),
\end{eqnarray}
which we hereafter write as $<fg>$.  From $ L = \Phi \Lambda \Psi$,
the entries of $ L$ are then expressed by
\begin{eqnarray}
\label{back}
a_{ij} := \left(  L\right)_{ij} =
 < \lambda \phi_{i} \psi_{j} > \ .
\end{eqnarray}
This gives a key equation for the inverse problem where we compute $ L$
from the eigenmatrix $\Phi$ (and $\Psi$) with the eigenvalues
$\lambda_i$.
So the eigenmatrix plays the role of the scattering data in the inverse
scattering method.
Then the method for solving the initial value
problem of the hierarchy (\ref{gtoda}) can be formulated as follows:
First we solve the eigenvalue (or scattering) problem (\ref{leq}) at $t=0$,
and find the scattering data, $\Phi^0:=\Phi(0)$.
Then we solve the time evolution of the eigenmatrix from (\ref{peq}),
and with the solution $\Phi(t)$
we obtain $ L(t)$ thorough the equation (\ref{back}).

\section{Inverse scattering method}
\renewcommand{\theequation}{3.\arabic{equation}}\setcounter{equation}{0}

In this section, we construct an explicit solution formula for the initial
value problem of the generalized Toda equation (\ref{gtoda}) by using
the inverse scattering method.  A key of this method is to generalize
the orthogonalization procedure of Szeg\"o with respect to the
inner product (\ref{product}).  This is essencially an extention of the method
developed in \cite{KM}.

\medskip

Following \cite{KM} we first ``gauge" transform
$ \Phi$ and $ \Psi$ by
\begin{eqnarray}
\label{trans}
 \Phi = G \tilde\Phi \ , \, \,  \Psi = \tilde\Psi G
\end{eqnarray}
where the matrix $G$ is given by
\begin{eqnarray}
\nonumber
G = diag \left[ < \tilde\phi_{1}\tilde\psi_{1}>^{-1/2}\ , \cdots \ , \
< \tilde\phi_{N}\tilde\psi_{N}>^{-1/2} \right] \ .
\end{eqnarray}
Note that the gauge transform (\ref{trans})
includes a freedom in the
choice of $\tilde\phi$ and $\tilde\psi$, that is, (\ref{trans}) is
invariant under
the scaling $\tilde\phi_{i} \ , \tilde\psi_i
\rightarrow f_{i}({\bf t}) \tilde\phi_{i}, \ f_i({\bf t})\tilde\psi_i$,
with $\{f_{i}\}_{i=1}^{N}$ arbitrary
functions of ${\bf t}$.  With (\ref{trans}), the equations
(\ref{leq}) and (\ref{peq}), as well as (\ref{dleq}) and (\ref{dpeq}),
 become
\begin{eqnarray}
\label{lam1}
& & \left( G^{-1}  L G \right) \tilde\Phi = \tilde\Phi \Lambda \ , \, \, \,
\tilde\Psi \left( G  L G^{-1} \right)  =  \Lambda \tilde\Psi \ , \\
\label{lam2}
& & \frac{d}{dt}\tilde\Phi =
\left( G^{-1}  P G \right) \tilde\Phi -
\left(\frac{d}{dt} \log{G} \right)\tilde\Phi \ , \\
& & \frac{d}{dt}\tilde\Psi =
- \ \tilde\Psi \left( G P G^{-1} \right)
- \ \tilde\Psi \left(\frac{d}{dt} \log{G} \right) \ .
\end{eqnarray}
Noting $G^{-1}(L)_{<0}G = (G^{-1} L G)_{<0}$ etc, we write
\begin{eqnarray}
\nonumber
 G^{-1} P G  &=&
 -2 \left( G^{-1} L G \right)_{<0} +
 G^{-1} L G  - diag\left( L \right) \ , \\
\nonumber
 G P G^{-1}  &=&
 2 \left( G  L G^{-1} \right)_{>0} -
 G  L G^{-1}  + diag\left( L \right) \ ,
\end{eqnarray}
from which we obtain the equations for the column vectors
$\tilde \phi(\lambda, t)$ in $\tilde \Phi$ and the row vectors
$\tilde \psi(\lambda, t)$ in $\tilde \Psi$,
\begin{eqnarray}
\label{nlam1}
\frac{ d \tilde\phi}{dt} &=&
 -2 \left( G^{-1} L G \right)_{<0} \tilde\phi +
\lambda \tilde\phi - \left( diag\left(  L \right) +
\frac{d}{dt} \log{ G } \right) \tilde\phi \ , \\
\label{nlam2}
\frac{d\tilde\psi}{dt} &=&
- 2\tilde\psi \left( G  L G^{-1} \right)_{>0}  +
\lambda \tilde\psi - \tilde\psi \left( diag\left( L \right) +
\frac{d}{dt} \log{ G } \right)  \ .
\end{eqnarray}
We here observe that (\ref{nlam1}) and (\ref{nlam2}) can be split into the
following sets of equations by fixing the gauge freedom in the determination
of $\phi$ and $\psi$. In the components, these are
\begin{eqnarray}
\label{cpn1}
& &\frac{ d \tilde\phi_i}{dt} =
- 2 \sum_{j=1}^{i-1} \frac{ < \lambda \tilde\phi_{i} \tilde\psi_{j} >}
{< \tilde\phi_j\tilde\psi_{j} >} \tilde\phi_{j} + \lambda \tilde\phi_{i} \ , \\
\label{cpm1}
& &\frac{ d\tilde\psi_j}{dt} =
- 2 \sum_{i=1}^{j-1} \tilde\psi_{i}\frac{ < \lambda\tilde\phi_{i} \tilde
\psi_{j} >}{<\tilde\phi_i\tilde\psi_{i} >}  + \lambda \tilde\psi_{j} \ , \\
\label{cpn2}
& &\frac{1}{2} \frac{ d}{dt} \log{
< \tilde\phi_i\tilde\psi_{i} > } =  a_{ii}  \ .
\end{eqnarray}
It is easy to check that (\ref{cpn1}) or (\ref{cpm1}) implies (\ref{cpn2}).
It is also immediate from (\ref{cpn1}) and (\ref{cpm1}) that we have:
\begin{Proposition}
The solutions of (\ref{cpn1}) and (\ref{cpm1}) can be written in the following
forms of separation of variables,
\begin{eqnarray}
\label{sepa1}
\tilde\phi(\lambda,t) &=& M(t) \phi^{0}(\lambda)
e^{\lambda t} \ , \\
\label{sepa2}
\tilde\psi(\lambda,t) &=& \psi^0(\lambda) N(t)
e^{\lambda t} \ ,
\end{eqnarray}
where $M(t)$ and $N(t)$ are, respectively, lower and upper
triangular matrices with $diag[M(t)] = diag[N(t)]
= I$, the identity matrix.
\end{Proposition}
\noindent
Note here that the initial data for $\tilde \phi$ and $\tilde \psi$ are chosen
as those of $\phi$ and $\psi$, i.e. $\tilde\phi(\lambda,0)=\phi(\lambda,0):=
\phi^0(\lambda)$ and $\tilde\psi(\lambda,0)=\psi(\lambda,0):=\psi^0(\lambda)$.
 As a direct consequence of this proposition, and the
orthogonality of the eigenvectors, (\ref{ortho2}), i.e. $< \tilde\phi_{i}
\tilde\psi_{j} > =0$ for $i \neq j$, we have:
\begin{Corollary}(Orthogonality):
For each $i,j \in \{2, \cdots, N\}$, we have for all
$t \in {\Bbb R}$,
\begin{eqnarray}
\label{corr1}
<\tilde\phi_{i} \psi_{\ell}^{0} e^{2\lambda t}> & = &0 \ , \, \, for \, \,
\ell = 1,2, \cdots , i-1 \, \ , \\
\label{corr2}
<\phi_{k}^0 \tilde\psi_{j} e^{2\lambda t }>  &=& 0 \ , \, \, for \, \,
k = 1,2, \cdots , j-1 \, .
\end{eqnarray}
\end{Corollary}
Now we obtain the formulae
for the eigenvectors of $L$ in terms of the initial data $\{
\phi_{i}^{0}(\lambda) \}_{1 \le i \le N}$ and
$\{\psi^0_j(\lambda)\}_{1 \le j \le N}$:
\begin{Theorem}
The solutions  $\tilde\phi_{i}(\lambda, t)$  and
$\tilde\psi_{j}(\lambda,t)$ of (\ref{cpn1}) and (\ref{cpm1}) are given by
\begin{eqnarray}
\label{phis}
\tilde\phi_{i}(\lambda,t) &=& \frac{e^{\lambda t}}
{ D_{i-1}(t)} \left|
\begin{array}{cccc}
c_{11} & \ldots & c_{1,i-1} & \phi_{1}^{0}(\lambda) \\
\vdots & \ddots & \vdots & \vdots \\
c_{i1} & \ldots & c_{i,i-1} & \phi_{i}^{0}(\lambda) \\
\end{array}
\right| \, , \\
\label{psis}
\tilde\psi_{j}(\lambda,t) &=& \frac{e^{\lambda t}}
{ D_{j-1}(t)} \left|
\begin{array}{ccc}
c_{11} & \ldots & c_{1j} \\
\vdots & \ddots & \vdots \\
c_{j-1,1} & \ldots & c_{j-1,j} \\
\psi_{1}^{0}(\lambda) & \ldots & \psi_{j}^{0}(\lambda) \\
\end{array}
\right|
\end{eqnarray}
where $c_{ij}(t) = < \phi_{i}^{0} \psi_{j}^{0} e^{ 2
\lambda t}>$, and $D_{k}(t)$ is the determinant of the
$k \times k$ matrix with entries $c_{ij}({\bf t})$, i.e.,
\begin{eqnarray}
\label{DDD}
D_{k}(t) = det\left[ \Big( c_{ij}(t) \Big)_{1 \le i,j \le k}
\right] \ .
\end{eqnarray}
(Note here that $c_{ij}(0)=\delta_{ij}$ and $D_k(0)=1$.)
\end{Theorem}
\begin{Proof}
{}From equation (\ref{corr1}) and (\ref{corr2}) with (\ref{sepa1})
and (\ref{sepa2}), we have
\begin{eqnarray}
\label{pro1}
\sum_{ \ell = 1}^{i} M_{i\ell}(t) < \phi_{\ell}^{0} \psi_{k}^{0} e^{2
\lambda t} > = 0 \ ,  \ \ for \ 1 \le k \le i-1 \ , \\
\label{pro2}
\sum_{ k = 1}^{j} < \phi_{\ell}^{0} \psi_{k}^{0} e^{2
\lambda t} >  N_{kj}(t) = 0 \ ,
\ \ for \ 1 \le \ell \le j-1 \ ,
\end{eqnarray}
Solving (\ref{pro1}) and (\ref{pro2}) for $M_{i\ell}$ and $N_{kj}$
with $M_{ii} = N_{jj} = 1$, we obtain
\begin{eqnarray}
M_{i\ell}(t) &=&\frac{(-1)^{i+\ell}}{D_{i-1}(t)} D_i \left[
\begin{array}{c}
\ell \\
i \\
\end{array}
\right] (t)
:=\frac{ \Delta_{\ell,i}({\bf t})}{D_{i-1}(t)} \ ,
\ \  1 \le \ell \le i \ , \\
N_{kj}(t) &= &\frac{(-1)^{k+j}}{D_{j-1}(t)} D_j \left[
\begin{array}{c}
j \\
k \\
\end{array}
\right] (t) :=
\frac{ \Delta_{j,k}({\bf t})}{D_{j-1}({\bf t})} \ ,
\ \ 1 \le  k \le j \ ,
\end{eqnarray}
where $
D_i \left[
\begin{array}{c}
\ell \\
i \\
\end{array}
\right]$
is the determinant of $D_i$ after removing the $\ell$-th row and $i$-th column.
 From (\ref{sepa1}) and (\ref{sepa2}), we then have
\begin{eqnarray}
\label{phi1}
& &\tilde\phi_{i}(\lambda, t) = e^{\lambda t}
\sum_{\ell=1}^{i} M_{i\ell}\phi_{\ell}^{0}
= \frac{e^{\lambda t}}{D_{i-1}(t)}
\sum_{\ell =1}^i \phi^0_{\ell}(\lambda)\Delta_{\ell, i}(t) \ , \\
\label{psi1}
& & \tilde\psi_{j}(\lambda ,t) = e^{\lambda t}
\sum_{k=1}^{j} \psi_k^{0}N_{kj}
= \frac{e^{\lambda t}}{D_{j-1}(t)} \sum_{k =1}^j
\Delta_{j, k}(t)\psi^0_k(\lambda) \ .
\end{eqnarray}
Noticing that $\Delta_{\ell, i}$ is the cofactor of the element $c_{\ell i}$
of the matrix $(c_{mn})_{1 \le m,n \le i}$ and $\Delta_{j,k}$ is for
$c_{jk}$ of
$(c_{mn})_{1 \le m,n \le j}$ we confirm (\ref{phi1}) and (\ref{psi1})
are just (\ref{phis}) and (\ref{psis}).
\end{Proof}

\medskip

\noindent
We then note:
\begin{Corollary}
The gauge factors $<\tilde\phi_{i}\tilde\psi_i>$ can be expressed by
\begin{eqnarray}
< \tilde\phi_{i}\tilde\psi_i >(t) = \frac{D_{i}(t)}{D_{i-1}(t)} \ .
\end{eqnarray}
\end{Corollary}
\begin{Proof}
{}From (\ref{phi1}) and (\ref{psi1}), we have
\begin{eqnarray}
\nonumber
\tilde\phi_i(\lambda)\tilde\psi_i(\lambda) = \frac{1}{D_{i-1}^2}
\sum_{\ell,k=1}^i \Delta_{\ell,i}\Delta_{i,k}
\phi_{\ell}^0(\lambda)\psi_k^0(\lambda) e^{2\lambda t} \ .
\end{eqnarray}
Then taking the bracket $<,>$ in (\ref{product}) leads to
\begin{eqnarray}
\nonumber
< \tilde\phi_i\tilde\psi_i> = \frac{1}{D_{i-1}^2} \sum_{\ell,k=1}^i
\Delta_{\ell,i}
\Delta_{i,k}c_{\ell k} \ .
\end{eqnarray}
Using the relation $\sum_{\ell =1}^i \Delta_{\ell,i}c_{\ell k} = D_i
\delta_{ik}$ with $\Delta_{i,i} = D_{i-1}$ complete the proof.
\end{Proof}

\medskip

\noindent
This yields the formulae for the normalized eigenfunctions
\begin{eqnarray}
\label{evcs1}
\phi_{i}(\lambda,t) &=& \frac{e^{\lambda t}}
{\sqrt{D_{i}(t)
D_{i-1}(t)}} \left|
\begin{array}{cccc}
c_{11} & \ldots & c_{1,i-1} & \phi_{1}^{0}(\lambda) \\
\vdots & \ddots & \vdots & \vdots \\
c_{i1} & \ldots & c_{i,i-1} & \phi_{i}^{0}(\lambda) \\
\end{array}
\right| \ \ , \\
\label{evcs2}
\psi_{j}(\lambda,t) &=& \frac{e^{\lambda t}}
{\sqrt{D_{j}(t)
D_{j-1}(t)}} \left|
\begin{array}{ccc}
c_{11} & \ldots & c_{1j} \\
\vdots & \ddots & \vdots \\
c_{j-1,1} & \ldots & c_{j-1,j} \\
\psi_{1}^{0}(\lambda) & \ldots & \psi_{j}^{0}(\lambda) \\
\end{array}
\right| \ \ .
\end{eqnarray}
With the formula (\ref{evcs1}) and (\ref{evcs2}),
 we now have the solution (\ref{back})
of the inverse scattering problem (\ref{leq}) and (\ref{peq}).

\medskip

The above derivation of the eigenvectors is the same as the orthogonalization
procedure of Szeg{\"{o}} \cite{Sz},
which is equivalent to the Gram - Schmidt
orthogonalization as observed in \cite{KM}.

\medskip
\noindent
{\bf Remark 1.}
The generalized Toda equation (\ref{gtoda}) with (\ref{pmx}) possesses a
hierarchy defined by
\begin{equation}
\label{htoda}
\frac{\partial}{\partial t_n} L = [P_n \ , \ L] \ , \, \, n = 1, 2, \cdots  \ ,
\end{equation}
where $P_n$ is given by
\begin{equation}
\label{pn}
P_n = \Pi(L^n) \equiv (L^n)_{>0} - (L^n)_{<0} \ .
\end{equation}
The commutativity of these flows can be shown by the ``zero" curvature
conditions of $P_n$, i.e.
\begin{equation}
\label{zeroc}
\frac{\partial P_m}{\partial t_n} - \frac{\partial P_n}{\partial t_m}
+ [P_m \ , \  P_n] = 0 \ ,
\end{equation}
which is a direct consequence of the choice of (\ref{pn}) \cite{KY}.
The solution for the hierarchy is then obtained by extending the argument
$\lambda t$ in the solution formula to $\xi (\lambda, t):=\sum_{n=1}^{\infty}
\lambda^n t_n \ $ \cite{KY}.

\medskip

\noindent
{\bf Remark 2.}
In \cite{N}, \cite{Sy} and  \cite{CN}, the authors  considered the following
generating matrix $P$ for the equation (\ref{gtoda}) with the general
matrix $L \in {\frak M}(N, {\Bbb R})$,
\begin{eqnarray}
\label{np}
P = (L)_{>0} - (L^T)_{<0} = (L)_{>0} - [(L)_{>0}]^T \ .
\end{eqnarray}
They then showed that this equation is integrable in the sense of a matrix
factorization of QR type, and the solution has the sorting property.
Our method developed in this section can be also applied to this problem
as follows:
First we note that the product $\Phi^{*}\Phi$
of the eigenmatrix $\Phi$ and its adjoint $\Phi^{*}:={\overline \Phi}^T$
is invariant under this flow (\ref{gtoda}).  Then we define a hermitian
matrix $S=(s_{ij})_{1 \le i,j \le N}$ as the inverse of ${\Phi}^{*}\Phi$,
i.e.
\begin{eqnarray}
\label{sortho}
\Phi S \Phi^{*} = I \ .
\end{eqnarray}
The matrix $S$ is determined from the initial eigenmatrix $\Phi^0$, and
$S \Phi^{*}$ gives the inverse of $\Phi$, that is, we have
$S \Phi^{*}$ for $\Psi$
in our method.  Note that if $L$ is symmetric, $S$ is an identity matrix $I$
and $\Phi \in O(N)$.  In general, we see from the Binet-Cauchy theorem that
$S$ is positive definite.  The equation (\ref{sortho}) now gives the
orthogonality relation,
\begin{eqnarray}
\label{csortho}
\sum_{1 \le k, \ell \le N} \phi_i(\lambda_k) s_{k\ell}
{\overline{\phi_j({\lambda}_{\ell})}} = \delta_{ij} \ ,
\end{eqnarray}
from which we define the following inner product as in (\ref{product}),
\begin{eqnarray}
\label{innerp}
\ll f , g \gg :=\sum_{1 \le k, \ell \le N} f(\lambda_k) s_{k\ell}
{\overline{g({\lambda}_{\ell})}} ={\overline{ \ll g , f \gg}} \ .
\end{eqnarray}
This leads to a positive definite metric.  Then following the procedure in
this section with some modifications based on $\Psi=S \Phi^{*}$,
we obtain the
same result for the eigenvectors (\ref{evcs1}) except the quantities
$c_{ij}$ which is now given by
\begin{eqnarray}
\label{scij}
c_{ij} = \ll \phi_i^0 e^{\lambda t} , \phi_j^0 e^{\lambda t} \gg
= {\overline {c}}_{ji} \ .
\end{eqnarray}
The solution $L(t)$ is then given by $L(t) = \Phi\Lambda S \Phi^{*}$, i.e.
\begin{eqnarray}
\label{saij}
a_{ij}(t) = \ll \lambda \phi_i , \phi_j \gg (t) \ .
\end{eqnarray}
Thus, we can show explicitly the integrability of the equation (\ref{gtoda})
with the generator $P$ given by (\ref{np}) for arbitrary diagonal matrix $L$,
and as a result of the positivity in the metric, the solution has the
sorting property.  The detail will be discussed elsewhere.

\section{Isospectral flows on simple Lie algebras}
\renewcommand{\theequation}{4.\arabic{equation}}\setcounter{equation}{0}

In this section, we consider the generalized Toda equations (\ref{gtoda})
associated with simple Lie algebras ${\frak g}$, and show their integrability.
 The matrices $L$ and $P$ here are given by a generalization of (\ref{bogl})
and (\ref{bogp}), i.e.
\begin{eqnarray}
\label{liel}
L_{\frak g}&=&\sum_{i=1}^r a_ih_i+\sum_{\alpha \in \Delta^{+}}
b_{\alpha}e_{\alpha}
+\sum_{\beta \in \Delta^{-}}c_{\beta}e_{\beta}, \\
\label{liep}
P_{\frak g}&=&\sum_{\alpha \in \Delta^{+}}b_{\alpha}e_{\alpha}
-\sum_{\beta \in \Delta^{-}}c_{\beta}e_{\beta} \ .
\end{eqnarray}
Here $h_i$ are the bases for the Cartan subalgebra with $r=rank({\frak g})$,
$\Delta^+$ and $\Delta^-$ are the sets of
positive and negative roots with the corresponding root
vectors $e_{\alpha}$ and $e_{\beta}(=e_{-\alpha})$.  These vectors $\{ h_i,
e_{\alpha} \}$ form the Cartan-Weyl bases of the simple Lie algebra
$\frak g$ which satisfy for $i,j \in \{1,\cdots,r\}$ and $\alpha, \beta \in
\Delta:=\Delta^+ \cup \Delta^-$
\begin{eqnarray}
\label{cwb}
\nonumber
& & [h_i , h_j] = 0, \, \, [h_i, e_{\alpha}] = \alpha(h_i)e_{\alpha} \ , \\
& & [e_{\alpha} , e_{\beta}] = N_{\alpha\beta}e_{\alpha+\beta}, \
{\hbox {if }}  \ \alpha+\beta
\in \Delta \ , \\
\nonumber
& & [e_{\alpha}, e_{-\alpha}] = h_{\alpha} , \ \  {\hbox {for }} \
\alpha \in \Delta^+ \ .
\end{eqnarray}
Using representations of the Cartan-Weyl bases, we now express (\ref{liel}) and
(\ref{liep}) in matrix form for each simple Lie algebra.  Then we prove that
the equation (\ref{gtoda}) with those $L_{\frak g}$ and $P_{\frak g}$
associated with the Lie algebra $\frak g$ is completely integrable by the
inverse
scattering method developed in Section 3.  The key ingredient in the proof
is to show that for each simple Lie algebra ${\frak g}$
there exists a ``permutation"
matrix $O_{\frak g}$ such that the matrices $L_{\frak g}$ and $P_{\frak g}$
are similar to $L$ and $P$ in (\ref{gtoda}) with $P$ defined by (\ref{pmx}),
i.e.
\begin{eqnarray}
\label{ol}
L &=& O_{\frak g} L_{\frak g} O_{\frak g}^T \ , \\
\label{op}
P &=& O_{\frak g} P_{\frak g} O_{\frak g}^T = \Pi (L) \ .
\end{eqnarray}
In another word, we look for a similarity transform such that the matrix
representations of $e_{\alpha}$ for $\alpha \in \Delta^+$ and $e_{\beta}$
for $\beta \in \Delta^-$ are transformed to strictly upper and lower
triangular matrices, respectively.  The existence of such representations is
due to Lie's theorem \cite{H}.  Here we give its explicit formula for
each semisimple Lie algebra.  Then the result in Section 3 implies
the integrability of the system (\ref{gtoda}) with $L_{\frak g}$ and
$P_{\frak g}$.  Note here that the generalized Toda equation is invariant
under the similarity transform. Here we consider all the classical simple
Lie algebras, $A_n, B_n, C_n$ and $D_n$.  The system associated
with the exceptional
algebra can be treated as the same way.  For convenient matrix representations
of the Cartan-Weyl bases, we follow the notations in \cite{C} and \cite{H}.

\medskip

\noindent
${\it A_{n}}$ :  Let $E_{ij}$ be the $(n+1) \times (n+1)$ matrix defined
in (\ref{e}).
We then take an element of the Cartan subalgebra as $h = \sum_{i=1}^{n+1}
\lambda_i E_{ii}$ with $\sum_{i=1}^{n+1}\lambda_i =0$.  Using (\ref{comm}) for
$E_{ij}$, we have
\begin{equation}
\label{ahe}
[ h , E_{ij} ] = (\lambda_i - \lambda_j) E_{ij} \ .
\end{equation}
Thus $E_{ij}$ gives a root vector corresponding to the root $\alpha(h)
=\lambda_i - \lambda_j$.  The simple roots are defined as
\begin{eqnarray}
\label{asrt}
\alpha_k (h)=\lambda_k-\lambda_{k+1}, \ {\hbox {for }}  \ k=1,
\cdots, n.
\end{eqnarray}
Then the positive (negative) roots are given by $\lambda_i - \lambda_j$
with $i<j$ $(i>j)$.  This implies that the choice of the $P_{A_{n}}$ is
the same as that in (\ref{pmx}).  Note also that adding some constant to
the Cartan subalgebra, one can choose $h_i$ of the basis to be $E_{ii}$.
Namely, the generalized Toda equation (\ref{gtoda}) with (\ref{l}) and
(\ref{p}) can be considered as an iso-spectral
flow on the simple Lie algebra $A_n$.

\medskip

\noindent
${\it C_m}$ : The element of this algebra is given by a $2m \times 2m$ matrix
$X$ satisfying $X^T J + J X =0$ where $J$ is defined by
\begin{eqnarray}
\label{j}
J=\left(
\begin{array}{cc}
0_m & I_m \\
-I_m & 0_m \\
\end{array}
\right).
\end{eqnarray}
Here $0_m$ is the $m \times m$ $0$-matrix, and $I_m$ is the $m \times m$
identity matrix.  We then choose the following bases with the $2m \times 2m$
matrix $E_{ij}$ defined in (\ref{e}),
\begin{eqnarray}
\nonumber
e_{ij}^1&=&E_{ij}-E_{j+m,i+m}, \ \ 1 \le i , j \le m \ ,\\
\label{ceij}
e_{ij}^2&=&E_{i,j+m}+E_{j,i+m}, \ \ 1 \le i\le j \le m \ ,\\
\nonumber
e_{ij}^3&=&E_{i+m,j}+E_{j+m,i}, \ \ 1 \le i\le j \le m \ .
\end{eqnarray}
Writing  $h = \sum_{i=1}^m \lambda_i e_{ii}^1$ as a general element of
the Cartan subalgebra, we have
\begin{eqnarray}
\nonumber
[h,e_{ij}^1]&=&(\lambda_i-\lambda_j)e_{ij}^1,\ \ i\neq j \ , \\
\label{cheij}
[h,e_{ij}^2]&=&(\lambda_i+\lambda_j)e_{ij}^2, \ \ i\le j \ ,\\
\nonumber
[h,e_{ij}^3]&=&-(\lambda_i+\lambda_j)e_{ij}^3, \ \ i\le j \ .
\end{eqnarray}
The simple roots are taken by
\begin{eqnarray}
\label{csrt}
\alpha_k(h)&=&\lambda_k-\lambda_{k+1}, \  \ {\hbox {for }}
\ 1 \le k \le m-1  \ , \\
\nonumber
\alpha_m(h)&=&2\lambda_m \ ,
\end{eqnarray}
from which the sets of positive and negative root vectors $\Sigma_{C_m}^+$ and
$\Sigma_{C_m}^-$ are given by
\begin{eqnarray}
\label{crtsp}
\Sigma_{C_m}^+ &=& \{ \ e_{ij}^1 , e_{k\ell}^2 \ | \ 1 \le i < j \le m , \
1 \le k \le \ell \le m \} \ , \\
\label{crtspm}
\Sigma_{C_m}^- &=& \{ \ e_{ij}^1 , e_{k\ell}^3 \ | \ 1 \le j < i \le m , \
1 \le k \le \ell \le m \} \ .
\end{eqnarray}
Then the matrix $L_{C_m}$ can be represented by
\begin{eqnarray}
\label{cml}
L_{C_m}=\left(
\begin{array}{cc}
A_1 & A_2 \\
A_3 & A_4 \\
\end{array}
\right) \ ,
\end{eqnarray}
where $A_1, \cdots, A_4$ are the $m \times m$ matrices satisfying the relations
\begin{eqnarray}
\label{crestriction}
A_1^T=-A_4,\ \  A_2=A_2^T, \ \ A_3=A_3^T \ .
\end{eqnarray}
The matrix $P_{C_m}$ is now given by
\begin{eqnarray}
\label{cmp}
P_{C_m}=\left(
\begin{array}{cc}
\Pi(A_1) & A_2 \\
- A_3 & -\Pi(A_4) \\
\end{array}
\right) \ .
\end{eqnarray}
We then obtain:
\begin{Proposition}
With the permutation matrix $O_{C_m}$, we have the generalized Toda equation
(\ref{gtoda}) on $C_m$ with $L$-$P$ pair given by
\begin{eqnarray}
\label{gcml}
L &=& O_{C_m} L_{C_m} O_{C_m}^T \ , \\
\label{gcmp}
P &=& O_{C_m} P_{C_m} O_{C_m}^T \ = \ \Pi(L) \ ,
\end{eqnarray}
where $O_{C_m}$ is given by
\begin{eqnarray}
\label{cmo}
O_{C_m}=\left(
\begin{array}{cc}
I_m & 0_m \\
0_m & Q_m \\
\end{array}
\right) \ ,
\end{eqnarray}
with the $m \times m$ matrix $Q_m$
\begin{eqnarray}
\label{qm}
Q_m=\left(
\begin{array}{llll}
0 &  \ldots & 0 & 1 \\
0 &  \ldots & 1 & 0 \\
0 & \vdots  & \vdots & 0 \\
1 &  \ldots & 0  & 0 \\
\end{array}
\right) = Q_m^T \ .
\end{eqnarray}
\end{Proposition}
\begin{Proof}
{}From (\ref{cml}) and (\ref{cmp}), it suffices to show
\begin{equation}
\label{a4}
-Q_m\Pi(A_4)Q_m^T = \Pi(Q_mA_4Q_m^T) \ .
\end{equation}
Recall that the multiplication of $Q_m$ from the left (right) implies
an exchange of rows (columns).  Then we see
\begin{equation}
\label{pa4}
Q_m(A_4)_{>0(<0)}Q_m^T = (Q_mA_4Q_m^T)_{<0(>0)} \ ,
\end{equation}
which implies the assertion.
\end{Proof}

\medskip
\noindent
Note that the equation (\ref{gtoda}) with $L_{C_m}$ and $P_{C_m}$
is just a reduction of the generalized Toda  equation on $A_{2m-1}$.

\medskip

\noindent
{\it Example 1}: We take the simplest case $C_2$.
The matrices $L_{C_2}$ and $P_{C_2}$ are represented as
\begin{eqnarray}
\label{C2l}
L_{C_2} = \left(
\begin{array}{llll}
a_1 & b_1 & b_2 & b_4 \\
c_1 & a_2 & b_4 & b_3 \\
c_2 & c_4 & -a_1 & -c_1 \\
c_4 & c_3 & -b_1 & -a_2 \\
\end{array}
\right)  ,
\end{eqnarray}
and
\begin{eqnarray}
\label{C2p}
P_{C_2} = \left(
\begin{array}{llll}
0 & b_1 & b_2 & b_4 \\
-c_1 & 0 & b_4 & b_3 \\
-c_2 & -c_4 & 0 & c_1 \\
-c_4 & -c_3 & -b_1 & 0 \\
\end{array}
\right).
\end{eqnarray}
Under the similarity transformation with $O_{C_2}$ defined in (\ref{cmo}),
$L_{C_2}$ and $P_{C_2}$ becomes
\begin{eqnarray}
\label{C2ll}
L = O_{C_2} L_{C_2}O_{C_2}^T= \left(
\begin{array}{llll}
a_1 & b_1 & b_4 & b_2 \\
c_1 & a_2 & b_3 & b_4 \\
c_4 & c_3 & -a_2 & -b_1 \\
c_2 & c_4 & -c_1 & -a_1 \\
\end{array}
\right),
\end{eqnarray}
and
\begin{eqnarray}
\label{C2pp}
P =  O_{C_2} P_{C_2}O_{C_2}^T = \left(
\begin{array}{llll}
a_1 & b_1 & b_4 & b_2 \\
-c_1 & a_2 & b_3 & b_4 \\
-c_4 & -c_3 & -a_2 & -b_1 \\
-c_2 & -c_4 & c_1 & -a_1 \\
\end{array}
\right).
\end{eqnarray}
Note here that under the similarity transformation the root space is decomposed
into the diagonal, upper and lower triangular parts of the matrix
(Lie's theorem).

\medskip

\noindent
${\it D_m}$:  The matrix representation of this algebra is given by a $2m
\times 2m$ matrix $X$ satisfying $X^TK + KX=0$, where $K$ is defined by
\begin{eqnarray}
\label{k}
K=\left(
\begin{array}{cc}
0_m & I_m \\
I_m & 0_m \\
\end{array}
\right).
\end{eqnarray}
Then the bases can be chosen as
\begin{eqnarray}
\nonumber
e_{ij}^1&=&E_{ij}-E_{j+m,i+m}, \ \ 1 \le i,j \le m \ ,\\
\label{deij}
e_{ij}^2&=&E_{i,j+m}-E_{j,i+m},  \ \ 1 \le  i < j \le m \ ,\\
\nonumber
e_{ij}^3&=&E_{i+m,j}-E_{j+m,i}, \ \  1 \le i < j \le m \ .
\end{eqnarray}
With a general element $h = \sum_{i=1}^{m} \lambda_i e_{ii}^1$ of the
Cartan subalgebra, we have
\begin{eqnarray}
\nonumber
[h,e_{ij}^1]&=&(\lambda_i-\lambda_j)e_{ij}^1, \ \ j\neq k \ ,\\
\label{dheij}
[h,e_{ij}^2]&=&(\lambda_i+\lambda_j)e_{ij}^2, \ \ i< j \ ,\\
\nonumber
[h,e_{ij}^3]&=&-(\lambda_i+\lambda_j)e_{ij}^3, \ \ i< j \ ,
\end{eqnarray}
from which the simple roots may be taken as
\begin{eqnarray}
\label{dsimple}
\alpha_k(h)&=&\lambda_k-\lambda_{k+1}, \ {\hbox {for }}
\ 1 \le k \le m-1  \ ,\\
\nonumber
\alpha_m(h)&=&\lambda_{m-1}+\lambda_m.
\end{eqnarray}
The sets of positive and negative root vectors $\Sigma_{D_m}^+$ and
$\Sigma_{D_m}^-$ are then given by
\begin{eqnarray}
\label{drtspp}
\Sigma_{D_m}^+ =\{ \  e_{ij}^1 , e_{ij}^2 \ | \ 1 \le i < j \le m \} \ , \\
\label{drtspm}
\Sigma_{D_m}^- =\{ \ e_{ij}^1 , e_{ji}^3 \ | \ 1 \le j < i \le m \} \ ,
\end{eqnarray}
The matrix $L_{D_m}$ is then expressed as
\begin{eqnarray}
\label{dml}
L_{D_m}=\left(
\begin{array}{cc}
A_1 & A_2 \\
A_3 & A_4 \\
\end{array}
\right) ,
\end{eqnarray}
where the $m \times m$ matrices $A_1, \cdots, A_4$ satisfy
\begin{eqnarray}
\label{drestriction}
A_1^T=-A_4,\ \  A_2=-A_2^T, \ \ A_3=-A_3^T.
\end{eqnarray}
The matrix $P_{D_m}$ is
\begin{eqnarray}
\label{dmp}
P_{D_m}=\left(
\begin{array}{cc}
\Pi(A_1) & A_2 \\
- A_3 & -\Pi(A_4) \\
\end{array}
\right) \ .
\end{eqnarray}
It is then immediate to see from Proposition 2 that the permutation matrix
$O_{D_m}$ is the same as in the case of $C_m$.  Namely we have:
\begin{Proposition}
With the permutation matrix $O_{D_m}=O_{C_m}$ given in (\ref{cmo}), we have
\begin{eqnarray}
\label{dl}
L&=&O_{D_m}L_{D_m}O_{D_m}^T \ , \\
\label{dp}
P&=&O_{D_m}P_{D_m}O_{D_m}^T = \Pi(L) \ .\
\end{eqnarray}
\end{Proposition}
\medskip

\noindent
${\it B_m}$ :
The element of this algebra satisfies the same relation as in $D_m$,
$X^TK + KX=0$, except now $K$ is the $(2m+1) \times (2m+1)$ matrix defined by
\begin{eqnarray}
\label{bk}
K=\left(
\begin{array}{ccc}
1 & {\pmb 0}^T & {\pmb 0}^T \\
{\pmb 0} & 0_m & I_m \\
{\pmb 0} & I_m & 0_m \\
\end{array}
\right),
\end{eqnarray}
where ${\pmb 0}$ is the $m$-column vector with $0$ entries.  This algebra
is referred as the orthogonal algebra $so(2m+1)$, while the algebra $D_m$
is as $so(2m)$, and has the same bases as (\ref{deij}) with additional
elements,
\begin{eqnarray}
\label{brootvs}
\nonumber
e_i^4&=&E_{0i}-E_{i+m,0} \ ,  \  1 \le i \le m \ , \\
e_i^5&=&E_{i0}-E_{0,i+m} \ , \ 1 \le i \le m \  ,
\end{eqnarray}
where we have labeled the indices of the matrix $E_{ij}$ as
$0 \le i,j \le 2m$.  With the expression $h = \sum_{i=1}^m \lambda_i e_{ii}^1$
as in the case of $D_m$, we have
\begin{eqnarray}
\label{broots}
[h,e_i^4]=-\lambda_ie_i^4,\ \ [h,e_i^5]=\lambda_ie_i^5 .
\end{eqnarray}
The simple roots are then chosen as
\begin{eqnarray}
\label{bsimple}
\alpha_k(h)&=&\lambda_k-\lambda_{k+1}, \ {\hbox {for }} \ 1 \le k \le m-1 \ ,
\\
\nonumber
\alpha_m(h)&=&\lambda_m \ .
\end{eqnarray}
The sets of positive and negative root vectors are now
\begin{eqnarray}
\label{brtsp}
\Sigma_{B_m}^+ &=& \{ \ e_{ij}^1 , e_{ij}^2 , e_k^5 \ | \
1 \le i < j \le m , \  1 \le k  \le m \} \ , \\
\label{brtspm}
\Sigma_{B_m}^- &=& \{ \ e_{ij}^1 , e_{ji}^3 , e_k^4 \ | \ 1 \le j < i
\le m , \ 1 \le k  \le m \} \ .
\end{eqnarray}
The matrix $L_{B_m}$ is then expressed as
\begin{eqnarray}
\label{bB}
L_{B_m}=\left(
\begin{array}{ccc}
0 & {\pmb b}_1^T & {\pmb b}_2^T \\
-{\pmb b}_2 & A_1 & A_2 \\
-{\pmb b}_1 & A_3 & A_4 \\
\end{array}
\right) \ ,
\end{eqnarray}
where ${\pmb b}_1, {\pmb b}_2$ are the $m$-column vectors, and the $m \times m$
matrices $A_1, \cdots, A_4$ satisfy the same relations as (\ref{drestriction}).
The matrix $P_{B_m}$ is now given by
\begin{eqnarray}
\label{bp}
P_{B_m}=\left(
\begin{array}{ccc}
0 & -{\pmb b}_1^T & {\pmb b}_2^T \\
-{\pmb b}_2 & \Pi(A_1) & A_2 \\
{\pmb b}_1 & -A_3 & -\Pi(A_4) \\
\end{array}
\right).
\end{eqnarray}
We then have:
\begin{Proposition}
With the $(2m+1) \times (2m+1)$ permutation matrix $O_{B_m}$, we have
\begin{eqnarray}
\label{bml}
L&=&O_{B_m}L_{B_m}O_{B_m}^T \ , \\
\label{bmp}
P&=&O_{B_m}P_{B_m}O_{B_m}^T = \Pi(L) \ .\
\end{eqnarray}
where $O_{B_m}$ is given by
\begin{eqnarray}
\label{bo}
O_{B_m}=\left(
\begin{array}{ccc}
{\pmb 0} & I_m & 0_m \\
1 & {\pmb 0}^T & {\pmb 0}^T \\
{\pmb 0} & 0_m & Q_m \\
\end{array}
\right).
\end{eqnarray}
\end{Proposition}
\begin{Proof}
Under the similarity transformation with $O_{B_m}$, $L$ and $P$ in (\ref{bml})
and (\ref{bmp}) are given by
\begin{eqnarray}
\label{olo}
L=\left(
\begin{array}{ccc}
A_1 & -{\pmb b}_2 & A_2Q_m \\
{\pmb b}_1^T & 0 & {\pmb b}_2^TQ_m \\
Q_mA_3 & -Q_m{\pmb b}_1 & Q_mA_4Q_m^T \\
\end{array}
\right),
\end{eqnarray}
and
\begin{eqnarray}
\label{opo}
P=\left(
\begin{array}{ccc}
\Pi(A_1) & -{\pmb b}_2 & A_2Q_m \\
-{\pmb b}_1^T & 0 & {\pmb b}_2^TQ_m \\
-Q_mA_3 & Q_m{\pmb b}_1 & -Q_m\Pi(A_4)Q_m^T \\
\end{array}
\right).
\end{eqnarray}
Then the equation (\ref{a4}) immediately leads to the result.
\end{Proof}

\medskip

\noindent
{\it Example 2}: We take the simplest case $B_2$, where $L_{B_2}$
and $P_{B_2}$ are represented by
\begin{eqnarray}
\label{B2l}
L_{B_2} = \left(
\begin{array}{lllll}
0 & c_3 & c_4 & -b_3 & -b_4 \\
b_3 & a_1 & b_1 & 0 & b_2 \\
b_4 & c_1 & a_2 & -b_2 & 0 \\
-c_3 & 0 & -c_2 & -a_1 & -c_1 \\
-c_4 & c_2 & 0 & -b_1 & -a_2 \\
\end{array}
\right),
\end{eqnarray}
and
\begin{eqnarray}
\label{B2p}
P_{B_2} = \left(
\begin{array}{lllll}
0 & -c_3 & -c_4 & -b_3 & -b_4 \\
b_3 & 0 & b_1 & 0 & b_2 \\
b_4 & c_1 & 0 & -b_2 & 0 \\
c_3 & 0 & c_2 & 0 & c_1 \\
c_4 & -c_2 & 0 & -b_1 & 0 \\
\end{array}
\right).
\end{eqnarray}
Under the similarity transformation with $O_{B_2}$ defined in (\ref{bo}), $L$
and $P$ in (\ref{olo}) and (\ref{opo}) become
\begin{eqnarray}
\label{B2ll}
L = \left(
\begin{array}{lllll}
a_1 & b_1 & b_3 & 0 & -b_2 \\
c_1 & a_2 & b_4 & b_2 & 0 \\
c_3 & c_4 & 0 & -b_4 & -b_3 \\
0 & c_2 & -c_4 & -a_2 & -b_1 \\
-c_2 & 0 & -c_3 & -c_1 & -a_1 \\
\end{array}
\right),
\end{eqnarray}
and
\begin{eqnarray}
\label{B2pp}
P = \left(
\begin{array}{lllll}
0 & b_1 & b_3 & 0 & -b_2 \\
-c_1 & 0 & b_4 & b_2 & 0 \\
-c_3 & -c_4 & 0 & -b_4 & -b_3 \\
0 & -c_2 & c_4 & 0 & -b_1 \\
c_2 & 0 & c_3 & c_1 & 0 \\
\end{array}
\right).
\end{eqnarray}

\section{Reductions on root spaces}
\renewcommand{\theequation}{5.\arabic{equation}}\setcounter{equation}{0}

As we have explained in the introduction, several generalizations of the
Toda equation may be obtained by taking reductions of the generalized Toda
equation (\ref{gtoda}) with general matrix $L$.  We then showed in the
previous section that the equations on simple Lie algebras studied in \cite{B}
are generalized by taking all the root vectors in the algebras.
In this section,
we consider reductions of these equations by restricting the set of
roots in the sums in (\ref{liel}).

\medskip

Let $S^+$ and $S^-$ be subsets of positive and
negative roots of a simple Lie algebra ${\frak g}$ defined by, for
$\forall \alpha_0 \in S^+$ and $\forall \beta_0 \in S^-$,
\begin{eqnarray}
\label{splus}
S^+ &:= &\{ \ \alpha \in \Delta^+ \ | \ \alpha \prec \alpha_0 \ \} , \\
\label{sminus}
S^- &:= &\{ \ \beta \in \Delta^- \ | \ \beta \succ \beta_0 \ \} .
\end{eqnarray}
Here ``$\prec$" and ``$\succ$" are the standard partial orderings between
roots.
We then consider the equation (\ref{gtoda}) with the matrices $\hat L$ and
$\hat P$ given by
\begin{eqnarray}
\label{lreduction}
\hat L&=&\sum_{i=1}^n a_ih_i+\sum_{\alpha\in S^+}b_{\alpha}e_{\alpha}
+\sum_{\beta\in S^-}c_{\beta}e_{\beta}, \\
\label{preduction}
\hat P&=&\sum_{\alpha\in S^+}b_{\alpha}e_{\alpha}
-\sum_{\beta\in S^-}c_{\beta}e_{\beta},
\end{eqnarray}
where $n=rank({\frak g})$.  We here claim:
\begin{Proposition}
\label{reduction}
The equation (\ref{gtoda}) with $\hat L$ and $\hat P$ is a
reduction of the generalized Toda equation on ${\frak g}$.
\end{Proposition}
\begin{Proof}
All we need to show is that the commutator $[\hat P, \hat L]$
is in the span of the root vectors whose roots are in $S^+$
and $S^-$.  From (\ref{lreduction}) and (\ref{preduction}),
 $[\hat L, \hat P]$ can be written as
\begin{eqnarray}
[\hat P, \hat L]= -\sum_{i=1}^n a_i [h_i,\hat P]+
2\sum_{\alpha\in S^+}\sum_{\beta\in S^-}b_{\alpha}c_{\beta}
[e_{\alpha},e_{\beta}]
\end{eqnarray}
Using (\ref{cwb}) we first note that the terms $[h_i, \hat P]$ does not
produce any new root vectors.  The second term, which then gives
$[e_{\alpha},e_{\beta}]=N_{\alpha\beta}e_{\alpha+\beta}$, has a root
 $\alpha + \beta
\in S^+ \cup S^-$ (if $ \alpha + \beta \in \Delta$),
since $\alpha \in \Delta^+$ and $\beta \in \Delta^-$.
This completes the proof.
\end{Proof}

\medskip

\noindent
{\it Example 3: The generalized Toda equation with band matrix $L$.}

\noindent
This example can be obtained as the following reduction on $A_{N-1}$:
Consider the subsets of the roots $S^+$ and $S^-$ given by
\begin{eqnarray}
\label{bandplus}
S^+ &=& \{ \ (i,j) \in \Delta^+ \ | \ 0 < j-i \le M^+ \le N-1 \} \ , \\
\label{bandminus}
S^- &=& \{ \ (i,j) \in \Delta^- \ | \ 0 < i-j \le M^- \le N-1 \} \ ,
\end{eqnarray}
where $M^+$ and $M^-$ are some positive integers.
Then the corresponding matrix $\hat L$ which we denote $L_{(M^+,M^-)}$
becomes
\begin{eqnarray}
\label{lhat}
L_{(M^+,M^-)}= \left(
\begin{array}{llllll}
a_{11} & \ldots & a_{1,1+M^+} & 0 & \ldots & 0 \\
\vdots & \ddots & \ldots & \ddots & \ldots & \vdots \\
a_{1+M^-,1} & \ldots & \ldots & \ldots & \ddots & 0 \\
0 & \ddots & \ldots & \ddots & \ldots & a_{N-M^+,N} \\
\vdots & \ldots & \ddots & \ldots  & \ddots & \vdots \\
0 & \ldots & 0 & a_{N,N-M^-} & \ldots  & a_{NN} \\
\end{array}
\right).
\end{eqnarray}

\medskip

As a special case of this example, we now costruct the full Kostant-Toda
equation having $L_H$-$P_H$ pair given in (\ref{hl}) and (\ref{hp}).  Here
we choose $S^+$ and $S^-$ to be the sets of the simple roots (i.e. $M^+=1$)
and of all the negative roots (i.e. $M^-=N-1$), respectively.
Thus the corresponding matrix is expressed as
\begin{eqnarray}
\label{hess}
L_{(1,N-1)} = \left(
\begin{array}{lllll}
a_{11} & b_1 & 0 & \ldots & 0 \\
a_{21} & a_{22} & b_2 & \ldots & \vdots \\
\vdots & \vdots & \ddots & \ddots & 0 \\
\vdots & \ldots & \ldots & a_{N-1,N-1} & b_{N-1} \\
a_{N1} & \ldots & \ldots & a_{N,N-1} & a_{NN} \\
\end{array}
\right) .
\end{eqnarray}
Then we claim:
\begin{Proposition}
The generalized Toda equation (\ref{gtoda}) with $L_{(1,N-1)}$ and
$P_{(1,N-1)} :=
\Pi(L_{(1,N-1)})$ is equivalent to the full Kostant-Toda equation having
$L_H$ and $P_H:=-2(L_H)_{<0}$ in (\ref{hl}) and (\ref{hp}).
\end{Proposition}
\begin{Proof}
Let $H$ be a matrix defined by
\begin{equation}
\label{hb}
H = diag \left[ \ 1, \ b_1,\ b_1b_2, \ \cdots\ ,
\prod_{i=1}^{N-1}b_i \ \right] \ .
\end{equation}
Then the matrix $L_H$ can be written as $L_H = HL_{(1,N-1)}H^{-1}$, which
corresponds to the rescaling of the eigenvectors, i.e. $\Phi_H = H\Phi$
with $L_H\Phi_H=\Phi_H\Lambda$.
{}From the equation (\ref{ctoda}) for $L_{(1,N-1)}$, we have
\begin{equation}
\label{beqs}
\frac{d b_i}{dt} = (a_{i+1,i+1} - a_{ii}) b_i , \ {\hbox {for }}
\ i=1, \cdots, N-1\ ,
\end{equation}
from which $H$ satisfies
\begin{equation}
\label{derh}
\frac{d}{dt} H= \Big( diag(L_H) - a_{11}I_N \Big) H \ .
\end{equation}
Note here that $diag(L_H) = diag(L_{(1,N-1)})= diag[a_{11}, \cdots, a_{NN}]$.
Then the derivative of $L_H$ is calculated as
\begin{equation}
\label{derhl}
\frac{d}{dt} L_H = \Big[diag(L_H)\ , \ L_H \Big] +
\Big[ \Pi(L_H) \ , \ L_H \Big] \ ,
\end{equation}
where we have used $dL_{(1,N-1)}/dt=[P_{(1,N-1)},L_{(1,N-1)}]$, and
$HP_{(1,N-1)}H^{-1} = \Pi(L_H)$.  Noting the relation
\begin{equation}
\label{pprel}
\Pi(L_H) + diag(L_H) = L_H - 2 (L_H)_{<0} \ ,
\end{equation}
we complete the proof.
\end{Proof}

\medskip
\noindent
Thus the full Kostant-Toda equation can be solved through the generalized
Toda equation with the $L_{(1,N-1)}$-$P_{(1,N-1)}$ pair as
the reduction on $A_{N-1}$, that is, with the solution $L_{(1,N-1)}$,
$L_H=HL_{(1,N-1)}H^{-1}$.

\section{Behaviors of the solutions}
\renewcommand{\theequation}{6.\arabic{equation}}\setcounter{equation}{0}

Here we study the behavior of the solution of the generalized Toda
equation obtained in Section 3 by following the approach in \cite{KY}.
Many results obtained in \cite{KY}  are valid for this more general situation.
First we note:
\begin{Lemma}
\label{real}
The determinants
$D_i$ for $ i = 1, 2, \cdots, N$ in (\ref{DDD}) are real functions.
\end{Lemma}
\begin{Proof}
In the construction of the solutions $\Phi(t)$ and  $\Psi(t)$,
the``gauge'' $G$ is fixed by (\ref{cpn2}).
In terms of $D_i$, (\ref{cpn2}) is
\begin{eqnarray}
\label{cpn3}
a_{ii}=\frac{1}{2} \frac{d}{dt} \log{
\frac{D_i}{D_{i-1}}}  .
\end{eqnarray}
Note that
$D_0 \equiv 1$, $D_i(0)=1$ and $a_{ii}$ are real functions. Then
we see by
induction that all $D_i$ are real functions.
\end{Proof}

\medskip
\noindent
In (\ref{cpn3}), suppose $D_i(t_0)=0$ for some finite $t_0$ and some $i$.
Then if $ L(t_0)$ is a finite matrix, $D_{i-1}(t_0)$ must be also 0.
By induction, $D_1(t_0)=0$, but $D_0(t) \equiv 1 $, this forces
$a_{11}$ to be infinite, which is a contradiction. So we have:
\begin{Lemma}
\label{zero}
Suppose $D_i(t_0)=0$ for some $t_0 < \infty$ and some $i$.  Then
$ L(t)$ blows up to infinity at $t_0$.
\end{Lemma}

We note that
$D_i$ for $ i = 1, 2, \cdots, N$  are the $i$-th leading principal minors
of the product of matrices $\Phi_e \Psi_e$, where
$\Phi_e$ and $\Psi_e$ are defined by
\begin{eqnarray*}
\Phi_e := \left(
\begin{array}{ccc}
e^{\lambda_1 t}\phi_{1}^{0}(\lambda_1) & \ldots &
e^{\lambda_N t}\phi_{1}^{0}(\lambda_N) \\
\vdots & \ddots & \vdots \\
e^{\lambda_1 t}\phi_{N}^{0}(\lambda_1) & \ldots &
e^{\lambda_N t}\phi_{N}^{0}(\lambda_N)
\end{array}
\right) ,
\end{eqnarray*}
and
\begin{eqnarray*}
\Psi_e &:=& \left(
\begin{array}{ccc}
e^{\lambda_1 t}\psi_{1}^{0}(\lambda_1) & \ldots &
e^{\lambda_1 t}\psi_{N}^{0}(\lambda_1) \\
\vdots & \ddots & \vdots \\
e^{\lambda_N t}\psi_{1}^{0}(\lambda_N) & \ldots &
e^{\lambda_N t}\psi_{N}^{0}(\lambda_N)
\end{array}
\right) .
\end{eqnarray*}
Then from the Binet-Cauchy theorem, we have:
\begin{Lemma}
\label{expansion}
The determinants
$D_i$ with $ i = 1, 2, \cdots, N$ can be expressed as
\begin{eqnarray}
\label{expand}
& & \\
\nonumber
& & D_i(t) = \sum_{J_{iN}}
e^{2\sum_{k=1}^i\lambda_{{j_k}}t} \left|
\begin{array}{ccc}
\phi_1^0(\lambda_{j_1}) & \ldots & \phi_1^0(\lambda_{j_i}) \\
\vdots & \ddots & \vdots \\
\phi_i^0(\lambda_{j_1}) & \ldots & \phi_i^0(\lambda_{j_i}) \\
\end{array}
\right|
\left|
\begin{array}{ccc}
\psi_1^0(\lambda_{j_1}) & \ldots & \psi_i^0(\lambda_{j_1}) \\
\vdots & \ddots & \vdots \\
\psi_1^0(\lambda_{j_i}) & \ldots & \psi_i^0(\lambda_{j_i}) \\
\end{array}
\right|,
\end{eqnarray}
where $J_{iN}=(j_1, \cdots, j_i)$ represents all possible combinations for
$1\le j_1<\cdots<j_i\le N$.
In particular $D_0(t) \equiv 1$, and $D_N(t)=\exp(2\sum_{i=1}^N\lambda_it)$.
\end{Lemma}
\noindent
This Lemma is very useful to study the asympototic behavior of
$D_i$ for large $t$.
\medskip

We now obtain:
\begin{Theorem}
Let the eigenvalues of $L$ be all real and ordered as
$\lambda_1>\lambda_2>\cdots>
\lambda_N$.
Suppose that $det (\Phi_{k}^{0}) \neq 0$ and $det(\Psi_k^0) \neq 0$
for $ k = 1, \ldots , N$, where $\Phi_{k}^{0}$ and $\Psi_k^0$ are
the $k$-th leading principal submatrices of $\Phi^{0}$ and $\Psi^0$,
respectively.
Then as $t \rightarrow \infty$, the eigenfunctions
$\phi_{i}(\lambda_{j},t)$ and $\psi_{j}(\lambda_{i},t)$ satisfy
\begin{eqnarray}
\label{asymptphi}
& & \phi_{i}(\lambda_{j},t) \rightarrow \delta_{ij} \times
\frac{ det (\Phi_{i}^{0}) det (\Psi_{i-1}^{0})}{\sqrt{det(\Phi_i^0\Psi_i^0)
det(\Phi_{i-1}^0\Psi_{i-1}^0)}}  \ , \\
\nonumber
& & \\
\label{asymptpsi}
& & \psi_{j}(\lambda_{i},t) \rightarrow \delta_{ij} \times
\frac{ det (\Phi_{i-1}^{0}) det (\Psi_{i}^{0})}{\sqrt{det(\Phi_i^0\Psi_i^0)
det(\Phi_{i-1}^0\Psi_{i-1}^0)}}  \ ,
\end{eqnarray}
which implies the sorting property as $t \rightarrow \infty$, that is,
$L(t)=\Phi(t)\Lambda\Psi(t) \rightarrow \Lambda$.
\end{Theorem}
\begin{Proof}
Here we give a proof for (\ref{asymptphi}).  The case for $\psi_j(\lambda_i,t)$
is obtained in the same way.
Using Lemma 3, and from the ordering in the eigenvalues we see that
the leading order term for $D_i$ is given by
\begin{eqnarray}
\label{leading}
D_i(t) \rightarrow e^{2 \sum_{k=1}^i\lambda_{{k}}t}
det(\Phi_i^0\Psi_i^0), \, \, {\hbox {as }} \, t \to \infty \ .
\end{eqnarray}
{}From (\ref{evcs1})
and (\ref{leading}), the eigenfunctions behave as $t \rightarrow \infty$
\begin{eqnarray}
\label{blt}
& & \\
\nonumber
& &\phi_{i}(\lambda; t) \rightarrow \frac{ e^{
\left( \lambda -2 \sum_{k=1}^{i-1} \lambda_{k} - \lambda_{i} \right) t}}
{\sqrt{det(\Phi_i^0\Psi_i^0) det(\Phi_{i-1}^0\Psi_{i-1}^0)} }  \left|
\begin{array}{cccc}
c_{11}(t) & \cdots & c_{1,i-1}(t) & \phi_{1}^{0}(\lambda) \\
\vdots & \ddots & \vdots & \vdots \\
c_{i1}(t) & \cdots & c_{i,i-1} & \phi_{i}^{0}(\lambda) \\
\end{array} \right| \ .
\end{eqnarray}
The dominant term in the determinant gives
\begin{eqnarray}
\label{dom}
& &
e^{2 \sum_{k=1}^{i-1} \lambda_{k}t} \sum_{{\Bbb P}_{i-1}}
\psi_{1}^{0}(\lambda_{\ell_{1}}) \cdots
\psi_{i-1}^{0}(\lambda_{\ell_{i-1}})
\left|
\begin{array}{cccc}
\phi_{1}^{0}(\lambda_{\ell_{1}}) & \ldots &
\phi_{1}^{0}(\lambda_{\ell_{i-1}}) & \phi_1^0(\lambda) \\
\vdots & \ddots & \vdots & \vdots \\
\phi_{i}^{0}(\lambda_{\ell_1}) & \ldots &
\phi_{i}^{0}(\lambda_{\ell_{i-1}}) & \phi_i^0(\lambda) \\
\end{array} \right| \\
\nonumber
& & \\
\nonumber
& & =
e^{2 \sum_{k=1}^{i-1} \lambda_{k}t} det(\Psi_{i-1}^0)
\left|
\begin{array}{cccc}
\phi_{1}^{0}(\lambda_{1}) & \ldots &
\phi_{1}^{0}(\lambda_{i-1}) & \phi_1^0(\lambda) \\
\vdots & \ddots & \vdots & \vdots \\
\phi_{i}^{0}(\lambda_{1}) & \ldots &
\phi_{i}^{0}(\lambda_{i-1}) & \phi_i^0(\lambda) \\
\end{array} \right| ,
\end{eqnarray}
where ${\Bbb P}_k$ is the permutation $\left(
\begin{array}{cccc}
1 & 2 & \cdots & k \\
\ell_1 & \ell_2 & \cdots & \ell_k \\
\end{array}
\right)$.  Noting that the
determinant in (\ref{dom}) is
zero for $\lambda = \lambda_{j}$, with $j=1, \ldots, i-1$,
we complete the proof.
\end{Proof}

\medskip

This theorem implies that if all the eigenvalues of $L$ are real, then generic
solutions have the ``sorting property'' in the asymptotic sense.
It should be however noted that $D_i(t)$ might be zero for some ``finite"
times, where the solution blows up (Lemma 2).  Next theorem provides sufficient
conditions for the solutions to blow up to infinity in finite time.

\begin{Theorem}
Suppose some eigenvalues of $L$ are not real,
$det\Phi_n^0\neq 0$ and  $det\Psi_n^0\neq 0$, for $n=1, \cdots, N$.
Then $L(t)$
blows up to infinity in finite time.
\end{Theorem}
\begin{Proof}
We order the eigenvalues of $\tilde L$ by their real parts. We still
assume all the eigenvalues to be distinct. Since $L$ is a real
matrix, the complex eigenvalues appear as pairs. For a convenience,
we also assume that there is at most one pair having the same real part.
Suppose $\lambda_k+i\beta_k$ and $\lambda_k-i\beta_k$ are the first pair
of complex eigenvalues. Then from  (\ref{expand}),
the leading order term in $D_k$ is
\begin{eqnarray*}
\label{complex}
e^{2 \sum_{l=1}^k\lambda_{{l}}t+ 2i\beta_kt} \left|
\begin{array}{ccc}
\phi_1^0(\lambda_{1}) & \ldots & \phi_1^0(\lambda_{k}+i\beta_k) \\
\vdots & \ddots & \vdots \\
\phi_k^0(\lambda_{1}) & \ldots & \phi_k^0(\lambda_{k}+i\beta_k) \\
\end{array}
\right|\left|
\begin{array}{ccc}
\psi_1^0(\lambda_{1}) & \ldots & \psi_1^0(\lambda_{k}+i\beta_k) \\
\vdots & \ddots & \vdots \\
\psi_k^0(\lambda_{1}) & \ldots & \psi_k^0(\lambda_{k}+i\beta_k) \\
\end{array}
\right| \\
\\
+ e^{2 \sum_{l=1}^k\lambda_{{l}}t- 2i\beta_kt} \left|
\begin{array}{ccc}
\phi_1^0(\lambda_{1}) & \ldots & \phi_1^0(\lambda_{k}-i\beta_k) \\
\vdots & \ddots & \vdots \\
\phi_k^0(\lambda_{1}) & \ldots & \phi_k^0(\lambda_{k}-i\beta_k) \\
\end{array}
\right|\left|
\begin{array}{ccc}
\psi_1^0(\lambda_{1}) & \ldots & \psi_1^0(\lambda_{k}-i\beta_k) \\
\vdots & \ddots & \vdots \\
\psi_k^0(\lambda_{1}) & \ldots & \psi_k^0(\lambda_{k}-i\beta_k) \\
\end{array}
\right| .
\end{eqnarray*}
Since $D_k$ is real by Lemma 1, one can write the above as
$$e^{2 \sum_{l=1}^k\lambda_{l}t}\Big[A\cos(2 \beta_kt)
+B\sin(2 \beta_kt)\Big].$$
where A and B are two real constants.
The above is an oscillating function about zero.
Thus by Lemma 2, we conclude that
$\tilde L(t)$ blows up to infinity in finite time.
\end{Proof}

\section{Example}
\renewcommand{\theequation}{7.\arabic{equation}}\setcounter{equation}{0}

In this section, we demonstrate the results obtained in this paper by taking
an explicit form of the matrix $L$.  The main purpose here
is to solve the generalized Toda equation (\ref{gtoda}) for this explicit
matrix, and discuss the behavior of the solution.

\medskip

Let us consider a $2 \times 2$ matrix $L(t)=(a_{ij})_{1 \le i,j \le 2}$.
The generalized Toda equation then gives
\begin{eqnarray}
\label{extoda}
\frac{d}{dt}\left(
\begin{array}{cc}
a_{11} & a_{12} \\
a_{21} & a_{22} \\
\end{array}
\right) = \left(
\begin{array}{cc}
2a_{12}a_{21} & a_{12}(a_{22} - a_{11}) \\
a_{21}(a_{22}-a_{11}) & -2a_{21}a_{12} \\
\end{array}
\right) .
\end{eqnarray}
The initial data of $L(t)$ is assumed to be
\begin{eqnarray}
\label{exl0}
L(0) = \left(
\begin{array}{cc}
0 & 1 \\
a & b \\
\end{array}
\right) ,
\end{eqnarray}
where $a$ and $b$ are arbitrary constatnts.
The eigenvalues of $L(0)$, $\lambda_1$ and $\lambda_2$, are
\begin{eqnarray}
\label{exevs}
\lambda_{1,2} = \frac{1}{2}\Big(b \pm \sqrt{b^2 +4a} \Big) .
\end{eqnarray}
Then the initial eigenmatrices $\Phi^0$ and $\Psi^0$ are expressed by
\begin{eqnarray}
\label{exphi0}
& &\Phi^0 = \left(
\begin{array}{cc}
1 & 1 \\
\lambda_1 & \lambda_2 \\
\end{array}
\right) , \\
\nonumber
\\
\label{expsi0}
& &\Psi^0 = \frac{1}{\lambda_2 - \lambda_1} \left(
\begin{array}{cc}
\lambda_2 & -1 \\
-\lambda_1 & 1 \\
\end{array}
\right) .
\end{eqnarray}
In order to compute the solutions $\Phi(t)$ and $\Psi(t)$ from (\ref{evcs1})
and (\ref{evcs2}), we need the quantities
$c_{ij}=<\phi^0\psi^0e^{2\lambda t}>$.  From (\ref{exphi0}) and (\ref{expsi0}),
they are
\begin{eqnarray}
\label{excij}
& & c_{11}(t) = \frac{1}{\lambda_2 - \lambda_1}\Big(
\lambda_2  e^{2\lambda_1 t} - \lambda_1 e^{2\lambda_2 t} \Big) , \\
\nonumber
& & c_{12}(t) = \frac{1}{\lambda_2 - \lambda_1}\Big(
- e^{2\lambda_1 t} + e^{2\lambda_2 t} \Big) , \\
\nonumber
& & c_{21}(t) = \frac{\lambda_1\lambda_2}{\lambda_2 - \lambda_1}\Big(
e^{2\lambda_1 t} - e^{2\lambda_2 t} \Big) , \\
\nonumber
& & c_{22}(t) = \frac{1}{\lambda_2 - \lambda_1}\Big(
-\lambda_1  e^{2\lambda_1 t} + \lambda_2 e^{2\lambda_2 t} \Big) ,
\end{eqnarray}
from which the determinants $D_i(t)$ in (\ref{DDD}) become
\begin{eqnarray}
\label{exddd}
 D_1(t) = c_{11}(t) , \,\, \
 D_2(t) = \left|
\begin{array}{cc}
c_{11}(t) & c_{12}(t) \\
c_{21}(t) & c_{22}(t) \\
\end{array}
\right| = e^{2(\lambda_1 + \lambda_2)t} .
\end{eqnarray}
We now have the solutions (Theorem 1),
\begin{eqnarray}
\label{exphisol}
\Phi(t) &=& \frac{1}{\sqrt{D_1(t)}} \left(
\begin{array}{cc}
e^{\lambda_1 t} & e^{\lambda_2 t} \\
\lambda_1 e^{\lambda_2 t} &  \lambda_2 e^{\lambda_1 t} \\
\end{array}
\right), \\
\nonumber
\\
\label{expsisol}
\Psi(t) &= &\frac{1}{(\lambda_2-\lambda_1)\sqrt{D_1(t)}} \left(
\begin{array}{cc}
\lambda_2 e^{\lambda_1 t} & - e^{\lambda_2 t} \\
-\lambda_1 e^{\lambda_2 t} &  e^{\lambda_1 t} \\
\end{array}
\right) .
\end{eqnarray}
The solution $L(t)$ of the generalized Toda equation is then obtained from
(\ref{back}), $a_{ij}(t) = <\lambda \phi_i\psi_j>(t)$,
\begin{eqnarray}
\label{exlsol}
\\
\nonumber
 L(t) = \frac{1}{\lambda_2  e^{2\lambda_1 t} - \lambda_1 e^{2\lambda_2 t}}
\left(
\begin{array}{cc}
{\lambda_1\lambda_2 \left(e^{2\lambda_1 t} - e^{2\lambda_2 t}\right)} &
{(\lambda_2 - \lambda_1)e^{(\lambda_1+\lambda_2) t}} \\
 & \\
 {-\lambda_1\lambda_2 (\lambda_2 - \lambda_1)e^{(\lambda_1+\lambda_2) t}} &
{\lambda_2^2 e^{2\lambda_1 t} - \lambda_1^2 e^{2\lambda_2 t}} \\
\end{array}
\right) .
\end{eqnarray}

\medskip

Now let us discuss the solution behavior for $t>0$.  First we assume both
eigenvalues $\lambda_1$ and $\lambda_2$ to be real.  With the choice of the
eigenvalues in (\ref{exevs}), we have $\lambda_1 \ge \lambda_2$.
Then if $\lambda_1\lambda_2 \le 0$, then the function $D_1(t)$ doesnot vanish
for all $t$.  This implies the sorting property (Theorem 2).
For the case of
$\lambda_1>\lambda_2>0$, the $D_1$ vanishes and we have
the blowing up in the solution at the time $t=t_B >0$,
\begin{eqnarray}
\label{tb}
t_B=\frac{1}{2(\lambda_1-\lambda_2)}\log \frac{\lambda_1}{\lambda_2}.
\end{eqnarray}
This formula also implies that for
$0>\lambda_1>\lambda_2$ we have the sorting result for $t>0$.
Note here that the blowing up
occurs at one time $t=t_B$ (\ref{tb}), and then the solution $L(t)$ will
be sorted as $t\rightarrow \infty$, with the asymptotic forms of the
eigenmatrices,
i.e. (\ref{asymptphi}) and (\ref{asymptpsi}),
\begin{eqnarray}
\label{asmphi}
& &\Phi (t) \rightarrow \sqrt{
\frac{\lambda_2-\lambda_1}{\lambda_2}}\left(
\begin{array}{cc}
1 & 0 \\
0 & \lambda_2 \\
\end{array}
\right). \\
\nonumber
\\
\label{asmpsi}
& &\Psi (t) \rightarrow \frac{1}{\sqrt{\lambda_2(
\lambda_2-\lambda_1)}}\left(
\begin{array}{cc}
\lambda_2 & 0 \\
0 & 1 \\
\end{array}
\right).
\end{eqnarray}

\medskip

For the case of the complex eigenvalue $\lambda_1={\bar \lambda_2}:=\alpha +
i\beta$, $D_1(t)$ is expressed as
\begin{eqnarray}
\label{d1t}
D_1(t)=e^{2\alpha t}\sec \theta \cos \left( 2\beta t+\theta \right)
\end{eqnarray}
with $\tan \theta={\alpha}/{\beta}$. This indicates the blowing up
(Theorem 3).

\medskip

In the case of  degenerate eigenvalues $\lambda_1=\lambda_2$ (i.e. $b^2+4a=0$),
we take the limit $\lambda_2 \to \lambda_1:=\lambda_0$ in (\ref{exlsol}), and
obtain
\begin{eqnarray}
\label{exldouble}
L(t)=\frac{1}{1- 2\lambda_0 t} \left(
\begin{array}{cc}
-2\lambda_0^2 t & 1 \\
 & \\
-\lambda_0^2 & 2\lambda_0(1-\lambda_0 t) \\
\end{array}
\right).
\end{eqnarray}
which showes the ``sorting property'' as $t\rightarrow \infty$, i.e.
$L(t) \rightarrow  \lambda_0 I_2$.
It should be noted however that $L(0)$ with the degenerate eigenvalues
 is not similar to the ``diagonal" matrix $\lambda_0 I_2$.

\medskip

We summarize the above results in Figure, where we classify the behavior
of the solution in terms of the parameters $a$ and $b$ in (\ref{exl0}).

\medskip

\centerline{Put Figure}

\medskip

\noindent
In the figure, the shaded region corresponds to the blowing up solutions for
$t>0$
where the eigenvalues are either complex or real with $\lambda_1>\lambda_2>0$.
The other region including the positive $b$-axis and the lower boundary of the
curve $b^2+4a=0$ gives the sorting property.

\par\medskip\medskip

\noindent
{\bf Acknowledgment}
The work of Y. Kodama is partially supported by an NSF grant DMS9403597.
J. Ye wishes to thank Prof. Guangtian Song for his encouragement through the
years.

%%%%%%%%%%%%%%%%%%%%%%%

\bibliographystyle{amsplain}

\end{document}